\documentclass{vgtc} 

\graphicspath{{figures/}{pictures/}{images/}{./}} 

\usepackage{times}                     

\usepackage{tabu}                      
\usepackage{booktabs}                  
\usepackage{lipsum}                    
\usepackage{mwe}                       

\usepackage{mathptmx}                  

\onlineid{0}

\vgtccategory{Research}

\vgtcinsertpkg

\title{Exploring Device-Oriented Video Encryption for Hierarchical Privacy Protection in AR Content Sharing}

\author{Yongquan Hu\thanks{e-mail: yongquan.hu@unsw.edu.au}\\ %
        \scriptsize University of New South Wales %
\and Dongsheng Zheng\thanks{e-mail: sam.zheng@student.unsw.edu.au}\\ %
     \scriptsize University of New South Wales %
\and Kexin Nie\thanks{e-mail: knie0519@uni.sydney.edu.au}\\ %
     \scriptsize University of Sydney %
\and Junyan Zhang\thanks{e-mail: junyanzhang0317@gmail.com}\\ %
     \scriptsize Tongji University %
\and Wen Hu\thanks{e-mail: wen.hu@unsw.edu.au}\\ %
     \scriptsize University of New South Wales %
\and Aaron Quigley\thanks{e-mail: aquigley@acm.org}\\ %
     \scriptsize CSIRO's Data61}

\teaser{
  \centering
  \includegraphics[width=0.95\linewidth]{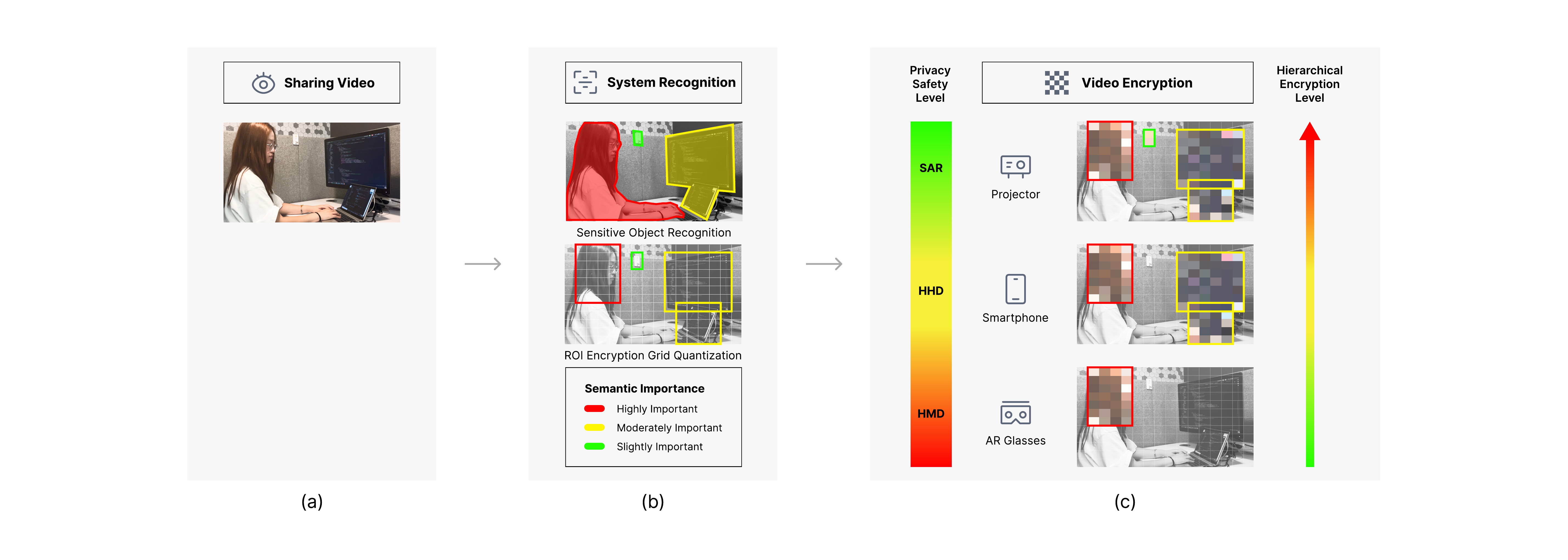}
  \caption{Preparation, execution, and results of the proposed method: (a) Preparing the shared AR video; (b) Objects in the video with varying semantic importance; (c) Considering different privacy levels for various AR displays and devices — Spatial Augmented Reality (SAR) projection, Handheld Display (HHD) smartphone, and Head Mounted Display (HMD) AR glasses — we implement device-oriented hierarchical encryption to balance privacy protection with real-time AR performance.}
  \label{fig:teaser}
}

\abstract{
Content sharing across multiple Augmented Reality (AR) displays is becoming commonplace, enhancing team communication and collaboration through devices like smartphones and AR glasses. However, this practice raises significant privacy concerns, especially concerning the physical environment visible in AR, which may include sensitive personal details like facial features and identifiable information. Our research focuses on protecting privacy within AR environments, particularly the physical backgrounds visible during content sharing across three common AR display methods: projection, smartphone, and AR glasses. We analyze the potential privacy risks associated with each method and employ a Region Of Interest (ROI) video encryption system to hierarchically encrypt the physical backdrop based on its safety rating. This study pioneers the integration of ROI video encryption at the bitstream level within AR contexts, providing a more efficient solution than traditional pixel-level encryption by enhancing encryption speed and reducing the required space. Our adaptive system dynamically adjusts the encryption intensity based on the AR display method, ensuring tailored privacy protection.
} 


\begin{document}


\firstsection{Introduction \& Related Work}\label{intro}

\maketitle

Augmented Reality (AR) bridges the virtual and real worlds, facilitating interactions across various domains \cite{nuernberger2016snaptoreality}. Investigating user experiences with various AR display technologies has been a central research theme. Recent studies have highlighted the benefits of integrating multiple AR displays \cite{hartmann2020aar} and promoting cross-AR display collaboration \cite{billinghurst2002collaborative} as promising avenues. Nevertheless, sharing identical content across different AR devices can pose privacy risks, particularly when sensitive elements from the physical environment, such as human faces, are involved \cite{cowan2021privacy}. Moreover, varying AR display methods entail differing risks of privacy breaches \cite{lehman2022hidden}. For instance, projection, a public display method, presents a lower security level due to its openness. In contrast, AR glasses, which are personal and viewed solely by the wearer, offer significantly higher privacy protection.

Visual encryption techniques, such as image or video encryption, are crucial for maintaining visual privacy \cite{padilla2015visual}. Research has been directed toward adapting these traditional encryption methods for XR scenarios. For instance, Du et al. developed an asymmetric image encryption algorithm that accommodates head-jitter, enabling different decryption results for public and private displays in Virtual Reality (VR) environments \cite{du2019tracking}. Comparatively, our research targets privacy protection issues specifically within AR contexts. We focus on the unintended exposure of real-world objects and scenes during content sharing—elements not inherent to the virtual overlay but part of the physical background. As illustrated in Figure \ref{fig:teaser}, sharing such scenes might inadvertently reveal personal information, such as faces and ID cards, which could discourage users from engaging. It is, therefore, crucial to apply layered privacy protections tailored to the specific AR display method used by the individual.

In this preliminary work, we initially explored device-oriented hierarchical encryption technology in the AR environment, focusing specifically on video encryption. We considered two critical factors in our system design: (1) encryption safety; (2) real-time performance of AR video. Theoretically, for security purposes, all sensitive objects in AR videos should be encrypted. However, studies indicate that the data scrambling inherent in video encryption can increase the computational complexity of the encoding-decoding process, inevitably causing video delays \cite{massoudi2008overview}. Meanwhile, real-time performance is vital for maintaining an intuitive AR user experience \cite{hu2023exploring}. These factors often oppose each other, necessitating a balance between them. Given that different sensitive objects in the visual field carry varying levels of semantic importance and, consequently, different privacy risks, we propose encrypting objects differently based on the AR display device used. This preliminary exploration suggests the feasibility of this approach, aiming to minimize data encryption to preserve real-time performance while maximizing privacy protection to ensure encryption security.

\section{Methodology}\label{methodology}
\subsection{ROI Video Encryption based on Bitstream Level}
\begin{figure}
\centering
\includegraphics[width=0.75\columnwidth]{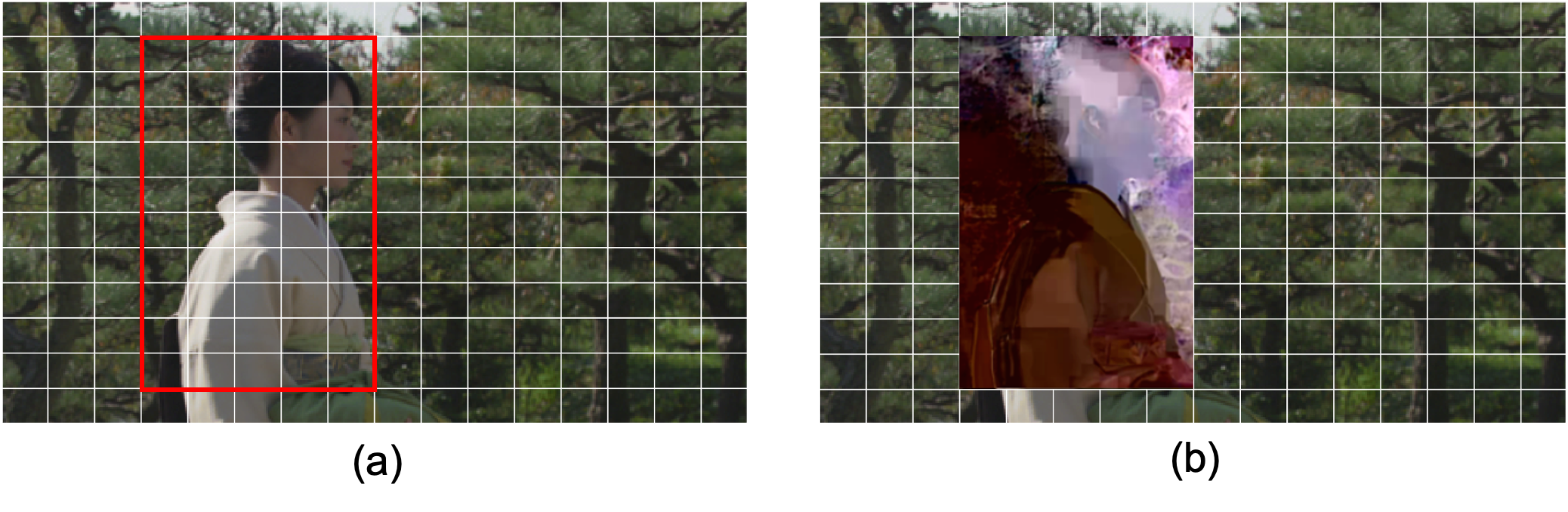}
  \caption{Tracking and encryption effect on Kimono1 standard test sequence based on ROSS system \cite{hu2018sdm} (video quantization parameters: 32; title number: 12$\times$16).}
  ~\label{fig:test}
\end{figure}

Region of Interest (ROI) video encryption integrates encryption directly into the visual encoding stage, enabling the system to automatically detect, or allowing users to manually select, specific areas or objects within the video for encryption \cite{farajallah2015roi}. This method conserves code streams and reduces encryption overhead. For better space optimization, encrypting at the code stream level — formed during compression — is more efficient than at the pixel level, delivering similar results with less key data. We utilize the  Region Of Semantic Saliency (ROSS) encryption system, as detailed in \cite{hu2018sdm}. ROSS, an ROI video encryption variant (as shown in Figure \ref{fig:test}), encrypts foreground objects algorithmically rather than through user selection. The High Efficiency Video Coding (HEVC) encoder, widely recognized in the field, supports ROI encryption via its ``tile'' mechanism, encoding distinct video areas independently. Notably, excessive ``tile'' can diminish encoding performance, impacting both speed and compression.

\subsection{AR Display Device Identification and Switching}
Extensive studies have investigated methods to seamlessly identify and transition between AR display devices \cite{suh2007context}. Common techniques involve utilizing device hardware details, software operating systems, or integrated sensors for differentiation. Currently, we manually determine the appropriate AR device for display and select the corresponding encryption level prior to each encryption.

\section{Findings \& Discussions}\label{findings}
Firstly, at a comparable encryption quality of PSNR$\approx$15dB, pixel-level encryption consumes 500MB, while the ROSS system at the bitstream level requires only 200KB. In our proposed AR hierarchical encryption, privacy protection levels vary by device, as demonstrated in Figure 1. It's crucial to note this is a mere example and not a thorough risk assessment. We speculate projectors, as public displays, pose the greatest privacy risk (i.e. the lowest privacy safety level). In contrast, smartphones offer users the ability to adjust sharing perspectives and choices, ensuring a balanced privacy stance. With AR glasses, which cater to a single viewer at once, users enjoy reduced risk of privacy breaches as the content is accessible to a limited number of people simultaneously. Additionally, as an example, we've simplified sensitive object categories to: human face (highly important), display content (moderately important), and ID cards (slightly important) in AR physical environment. In short, the proposed system aims to adjust the encryption level based on the associated privacy safety risk; more exposed devices will undergo increased encryption.

\section{Limitations, Future Work \& Conclusions}\label{conclusions}
The proposed method supports hierarchical encryption of specific objects, and we plan to expand the range of test objects in the future. Currently, the system only supports identifying and switching between AR display devices; however, we aim to develop a fully automated end-to-end system by integrating additional components. Future work will also include more comprehensive experiments, such as user studies and performance tests, to thoroughly evaluate both quantitatively and qualitatively the impact of this method on the two design factors mentioned earlier: encryption safety and AR video real-time performance.


\bibliographystyle{abbrv-doi}

\bibliography{template}
\end{document}